\documentclass[aps,twocolumn,showpacs]{revtex4}
\usepackage{graphicx}
\usepackage{bm}
\usepackage{amsmath}
\begin{document}

\title{Temperature dependence of Vortex Charges in High
Temperature Superconductors}

\author{Yan Chen,$^1$ Z. D. Wang,$^{2,3}$ and C. S. Ting$^1$}

\affiliation{$^1$Texas Center for Superconductivity and Department
of Physics, University of Houston, Houston, TX 77204\\
$^2$Department of Physics, University of Hong Kong, Pokfulam Road,
Hong Kong, China\\ $^3$Department of Material Science and
Engineering, University of Science and Technology of China, Hefei
230026, China}

\begin{abstract}

Using  a model Hamiltonian with $d$-wave superconductivity and
competing antiferromagnetic (AF) interactions, the temperature
($T$) dependence of the vortex charge in high $T_c$
superconductors is investigated by numerically solving  the
Bogoliubov-de Gennes equations. The strength of the induced AF
order inside the vortex core is $T$ dependent. The vortex charge
could be negative when  the AF order with sufficient strength is
present at low temperatures. At higher temperatures, the AF order
may be completely suppressed and the vortex charge becomes
positive. A first order like transition in the $T$ dependent
vortex charge is seen near the critical temperature $T_{AF}$. For
underdoped sample, the spatial profiles of the induced
spin-density wave and charge-density wave orders could have stripe
like structures at $T < T_s$, and  change  to two-dimensional
isotropic ones at $T > T_s$. As a result, a vortex charge
discontinuity occurs at $T_s$.

\end{abstract}


\maketitle

For the past many years there have been intensive studies of the
vortex physics  in high temperature superconductors (HTS). It is
well established now that the $d$-wave superconductivity (DSC)
with a competing antiferromagnetism (AF) order plays an important
role in determining the vortex structure of HTS. Many theoretical
studies~\cite{Arovas97,Ogata99,Demler,Machida,Zhu01,Zhang02,chen02,chen03,Franz,Ghosal}
have shown that the AF order may appear and coexist with the
underlying vortices. Experimental facts including neutron
scattering~\cite{Lake01}, moun spin rotation
measurement~\cite{Miller} and nuclear magnetic resonance (NMR)
experiments~\cite{NMR,NMR3} have provided a strong support for the
existence of AF order inside the vortex core in appropriately
doped HTS. On the other hand, the electrical charge associated
with the vortex in superconductors has also been paid considerable
attention both
theoretically~\cite{KF95,Blatter95,Machida98,KLB01,Brandt2,vcharge}
and experimentally~\cite{Matsuda98,Matsuda01}. Based on the BCS
theory, Blatter {\em et al.}~\cite{Blatter95} pointed out that for
$s$-wave superconductor the vortex charge is proportional to the
slope of the density of states at the Fermi level. However, the
NMR and nuclear quadrupole resonance (NQR) measurements on
YBa$_2$Cu$_3$O$_7$ and YBa$_2$Cu$_4$O$_8$~\cite{Matsuda01} seemed
to obtain results for the vortex charge, contradictory to that
predicted from the existing BCS theory with respect to both sign
and order of magnitude. In addition, if the impact  of vortex
charge on the mixed state Hall signal is considered~\cite{KF95},
the sign estimated from Hall effect experiments~\cite{Matsuda98}
for various HTS materials disagrees with the prediction of
BCS-type theory. In view of these points, together with the fact
that the strong electron correlation with the $d$-wave
superconducting pairing was not considered for the vortex charge,
we carried out the previous studies~\cite{vcharge} at zero
temperature and found that the vortex charge in HTS is strongly
influenced by the presence of the induced AF order in the vortex
core. The charge carried by a vortex is always positive for a pure
$d$-wave superconductor, and it  becomes  negative when there is
sufficient strength of AF order in the vortex core~\cite{vcharge}.
Since most relevant experimental phenomena have been observed at
finite $T$ not close to zero, including the sign change of the
mixed state Hall effect~\cite{Matsuda98},
it is necessary and valuable to study theoretically  the vortex
charges at finite $T$,  which so far has not been addressed in the
literature.

In this article we shall investigate the vortex charge as a
function of $T$ for both optimally doped and underdoped  HTS in
detail. Our calculation will be based on the model Hamiltonian on
a square lattice with a nearest neighboring  attractive
interaction $V_{DSC}$ describing the DSC and an onsite Coulomb
repulsion $U$ representing the competing AF order. We shall adopt
the well-developed numerical scheme~\cite{Wang95} to study the
vortex structure. Our numerical results at finite $T$ confirm that
the sign of the  vortex charge is still determined by the AF order
in the vortex core, being consistent with  our previous
work~\cite{vcharge}. The vortex charge could be transformed from
negative to positive values as $T$ is increased to a critical
value $T_{AF}$ such that  the AF order is completely suppressed at
$T > T_{AF}$, and the transition seems to be first-order-like. For
underdoped sample we show that the symmetry of the induced
spin-density wave (SDW) and the charge-density wave (CDW) may
change from stripe like to two dimensional like  at a critical
temperature $T_s$. A discontinuity of the $T$-dependent vortex
charge also appears at $T_s$.

The effective mean-field Hamiltonian describing both DSC and SDW
orders on a two-dimensional lattice can be represented as,
\begin{eqnarray}
H&=&-\sum_{i,j,\sigma} t_{ij} c_{i\sigma}^{\dagger}c_{j\sigma}
+\sum_{i,\sigma}( U n_{i {\bar {\sigma}}} -\mu)
c_{i\sigma}^{\dagger} c_{i\sigma} \nonumber \\ &&+\sum_{i,j} (
{\Delta_{ij}} c_{i\uparrow}^{\dagger} c_{j\downarrow}^{\dagger} +
h.c.)\;,
\end{eqnarray}
where $c_{i\sigma}^{\dagger}$ is the electron creation operator,
$\mu$ is the chemical potential. In the presence of magnetic field
$B$ perpendicular to the plane, the hopping integral can be
expressed as $t_{ij}= t_0 \exp[{i \frac{ \pi}{\Phi_{0}} \int_{{\bf
r}_{j}}^{{\bf r}_{i}} {\bf A}({\bf r})\cdot d{\bf r}}]$ for the
nearest neighboring sites $(i,j)$. A Landau gauge ${\bf
A}=(-By,0,0)$ is chosen. The two possible SDW and DSC orders in
cuprates are defined as $\Delta^{SDW}_{i} = U \langle c_{i
\uparrow}^{\dagger} c_{i \uparrow} -c_{i \downarrow}^{\dagger}c_{i
\downarrow} \rangle$ and $\Delta_{ij}=V_{DSC} \langle
c_{i\uparrow}c_{j\downarrow}-c_{i \downarrow} c_{j\uparrow}
\rangle /2$, where $U$ and $V_{DSC}$ represent respectively the
interaction strengths for two orders. The mean-field Hamiltonian
(1) can be diagonalized by solving the resulting Bogoliubov-de
Gennes equations self-consistently
\begin{equation}
\sum_{j} \left(\begin{array}{cc} {\cal H}_{ij,\sigma} &
\Delta_{ij} \\ \Delta_{ij}^{*} & -{\cal H}_{ij,\bar{\sigma}}^{*}
\end{array}
\right) \left(\begin{array}{c} u_{j,\sigma}^{n} \\
v_{j,\bar{\sigma}}^{n}
\end{array}
\right) =E_{n} \left(
\begin{array}{c}
u_{i,\sigma}^{n} \\ v_{i,\bar{\sigma}}^{n}
\end{array}
\right)\;,
\end{equation}
where the single particle Hamiltonian ${\cal H}_{ ij,\sigma}=
-t_{ij} +(U n_{i \bar{\sigma}} -\mu)\delta_{ij}$, and $n_{i
\uparrow} = \sum_{n} |u_{i\uparrow}^{n}|^2 f(E_{n})$, $ n_{i
\downarrow} = \sum_{n} |v_{i\downarrow}^{n}|^2 ( 1- f(E_{n}))$, $
\Delta_{ij} = \frac{V_{DSC}} {4} \sum_{n} (u_{i\uparrow}^{n}
v_{j\downarrow}^{n*} +v_{i\downarrow}^{*} u_{j\uparrow}^{n}) \tanh
\left( \frac{E_{n}} {2k_{B}T} \right)$, with $f(E)$ as the Fermi
distribution function. The DSC order parameter at the $i$th site
is $\Delta^{D}_{i}= (\Delta^{D}_{i+e_x,i} + \Delta^{D}_{ i-e_x,i}
- \Delta^{D}_{ i,i+e_y} -\Delta^{D}_{ i,i-e_y})/4$ where $
\Delta^{D}_{ij} = \Delta_{ij} \exp[ i { \frac{\pi}{\Phi_{0}}
\int_{{\bf r}_{i}}^{({\bf r}_{i}+{\bf r}_{j})/2 } {\bf A}({\bf r})
\cdot d{\bf r}}]$ and ${\bf e}_{x,y}$ denotes the unit vector
along $(x,y)$ direction. In our calculation, the related
parameters are chosen as: $t=a=1$, $V_{DSC}=1.2$, the linear
dimension of the unit cell of the vortex lattice is $N_x \times
N_y = 40 \times 20$. This choice corresponds the magnetic field $B
\simeq 37 T$.

In the absence of the magnetic field and for optimally doped
sample ($\delta=0.15$), the parameters are chosen in such a way
that only DSC (with $T_{c} \sim 0.164)$ prevails in the system.
Our numerical results with $U=2.2$ indeed show that the AF order
is induced at low temperature and is suppressed when $T$ is
larger.
\begin{figure}[b]
\includegraphics[width=8.6cm]{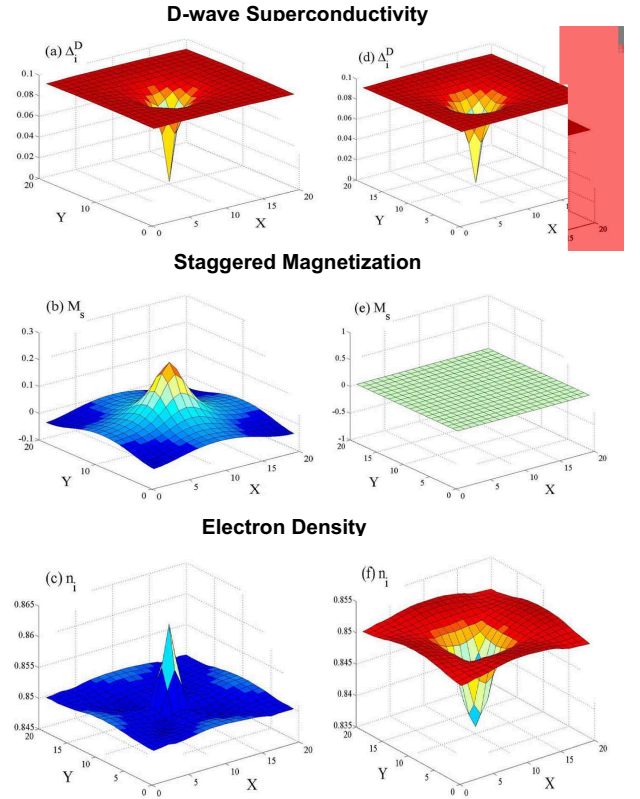}
\caption{\label{Fig1} Spatial variations of the DSC order
parameter $\Delta_{i}^{D}$ [(a) and (d)], staggered magnetization
$M_{i}^{s}$ [(b) and (e)], and net electron density $n_{i}$ [(c)
and (f)] in a $20 \times 20$ lattice. The left panels [(a), (b),
and (c)] and the right panels [(d), (e), and (f)] are for $T=0.02$
and $T=0.05$, respectively. The averaged electron density is fixed
at $\bar {n}=0.85$ and $U=2.2$.}
\end{figure}
The spatial profiles of the superconductivity order parameter,
staggered magnetization, and the charge density near the vortex
core are plotted at two different temperatures in Fig.1. Here the
staggered magnetization of the induced AF or SDW order is defined
as $M_{s}^{i} =(-1)^{i} \Delta^{SDW}_{i}/U$. The AF order is
generated in the region where the DSC order parameter is partially
suppressed. The low temperature results at $T =0.02$  are shown in
Figs. 1(a) to 1(c) where the AF order is nucleated and spreads out
from the core center, and the vortex charge is negative. The
induced AF order behaves like a two dimensional SDW with the same
wavelength in the $x$ and $y$ directions.  The higher temperature
results at $T=0.05$ are shown in Figs. 1(d) to 1(f) where the AF
order is completely suppressed and the vortex charge becomes
positive.  The appearance of the SDW order pinned by  the vortex
lattice at low $T$ strongly enhances the net electron density (or
depletion of the hole density) at the vortex core as shown in
Fig.1(c). The present result is consistent with that of Ref. 20
where the doping and $U$-value dependences of vortex charges are
examined. Both works reach the same conclusion: the vortex charge
is negative when a sufficient strength of SDW order is induced ,
while it turns to positive with the suppression of SDW order due
to increasing $T$ or  doping level, or decreasing the $U$-value.

Fig. 2(a) plots the phase diagram of $T$ versus $U$ for positively
(hole-rich) and negatively (electron-rich) charged vortices. It is
obvious that the AF vortex core can easily be induced in the low
temperature regime or with stronger AF interaction $U$ while the
pure DSC core tends to exist in the high $T$ or smaller $U$.  The
electron density inside the core is higher than the average
density in the low temperature region while the electron density
becomes lower than the average in the high temperature region.
There exists a clear boundary between these two phases. To
estimate the vortex charge, we first determine the vortex size by
examining the spatial profile of DSC order parameter. Next we sum
over the net electron density inside the vortex core.  In Fig.
2(b),  we depict the chemical potential versus $T$ at fixed doping
level ($\delta=0.15$ with $U=2.4$ and $U=2.2$). The result (solid
line) for $U=2.4$ exhibits a first-order like transition at
$T_{AF} \sim 0.077$, indicating the existence of induced AF order
below this critical temperature. For $U=2.2$, the result (dashed
line) seems to show a very weak discontinuity which can hardly be
seen in Fig. 2(b).
\begin{figure}[t]
\includegraphics[width=9.1cm]{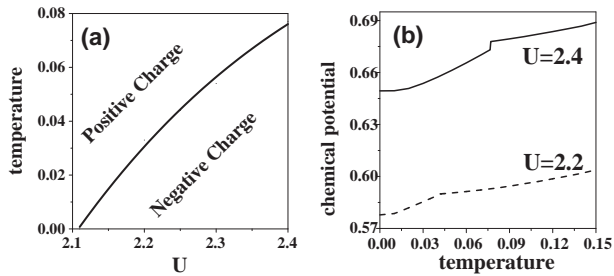}
\vspace{-0.9cm} \caption{\label{Fig2} (a) Phase diagram of
interaction strength $U$ versus temperature $T$ for positive and
negative charged vortex at optimal doping $\delta=0.15$. (b)
temperature dependence of chemical potential for $U=2.4$ and
$U=2.2$.}
\end{figure}
The above finding may be closely related to the experimental
results for underdoped TlBa$_2$Cu$_4$O$_8$~\cite{NMR3} where the
vortex core with  induced AF order below  $T_{AF}$  was reported.
Also from Fig. 2(b), a discontinuity in the slope of the
$T$-dependent chemical potential at $T_{AF}$ for both cases can be
clearly observed. We expect that the local density of states
(LDOS) of the vortex core should have double
peaks~\cite{Zhu01,chen02, Ghosal,Renner,Pan01} near zero bias for
$T <T_{AF}$; while for $T > T_{AF}$ only a single  broad zero-bias
peak can appear in the LDOS as one of the essential
characteristics of DSC~\cite{Wang95}. It would be interesting to
measure this temperature evolution in the LDOS by future STM
experiments. Also interestingly, even though the origin of Hall
sign anomaly in HTS is still debatable~\cite{Wang94}, the vortex
charge could make an additional contribution to the sign change in
the mixed state Hall conductivity~\cite{KF95}. As a result, we
predict that a sign reversal in the Hall signal may occur at $T$
not too far below $T_{AF}$.

In Fig. 3, we plot the  vortex charge number $Q_v/e$ as a function
of  $T$ for three typical $U$-values, here $e<0$ is the charge of
an electron.
\begin{figure}[t]
\includegraphics[width=7.2cm]{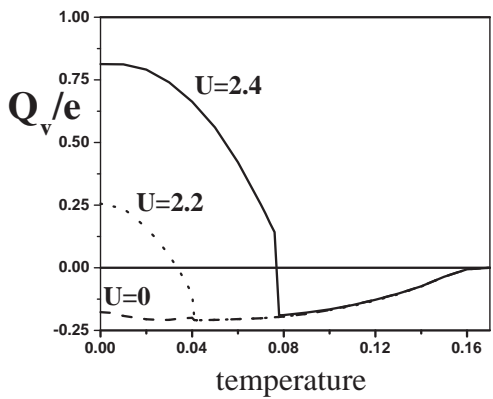}
\vspace{-0.5cm} \caption{\label{Fig3} The temperature dependence
of the number of vortex charge $Q_v/e$ for $U=2.4$(solid line),
$U=2.2$(dotted line) and $U=0$(dashed line). The averaged electron
density is fixed at $\bar {n}=0.85$.}
\end{figure}
The $T$ dependence of the induced staggered magnetization
$M_{s}^{c}$ at the vortex core center has also been examined  in
previous studies~\cite{Taki03,chenNMR} where  the induced AF order
is shown to disappear for $T > T_{AF}$.  From Fig. 3 and for
$U=2.4$, one can  clearly see an abrupt jump in the number of
vortex charge $Q_v/e$ as $T$ varies around $T_{AF} \sim 0.077$
which is consistent with the critical temperature as shown in Fig.
2(b). This positive to negative vortex charge transition seems  to
be first-order like. The magnitude of the discontinuity reduces to
one third when $U=2.2$ at $T_{AF}\sim 0.04$. Here the
discontinuity occurs at the positive to positive vortex charge
transition, not the positive to negative transition as studied in
the case of $U=2.4$. This is because slightly below $T_{AF}$, the
induced AF order is too weak to make the vortex charge negative.
For a pure DSC case with $U=0$, the vortex charge $Q_v$ is always
positive and its magnitude first increases slightly with $T$ and
then  decreases to zero as $T$ approaches to $T_c$. The critical
temperature $T_{AF}$ is $U$-value dependent or the
sample-dependent. The larger $U$ corresponds to larger $T_{AF}$.
These results confirm that the vortex charge is strongly
influenced by two competing effects: the suppression of the DSC
order at the core center leading to the depletion of the
electrons, and the induction of the AF order which favors the
accumulation of electrons. The condition for the negative vortex
charge to appear depends solely on whether there is a sufficient
AF order inside the vortex core. We emphasize that our results are
robust and indifference to band parameters, and should give a
qualitative description on the vortex physics in HTS.

In the following we are going to investigate the temperature
dependence of the vortex charge for underdoped HTS ($\delta
=0.12$). For $U=2.4$, the spatial profiles of the staggered
magnetization and electron density near the vortex core  at three
different temperatures are plotted in Fig. 4.
\begin{figure}[t]
\includegraphics[width=8.5cm]{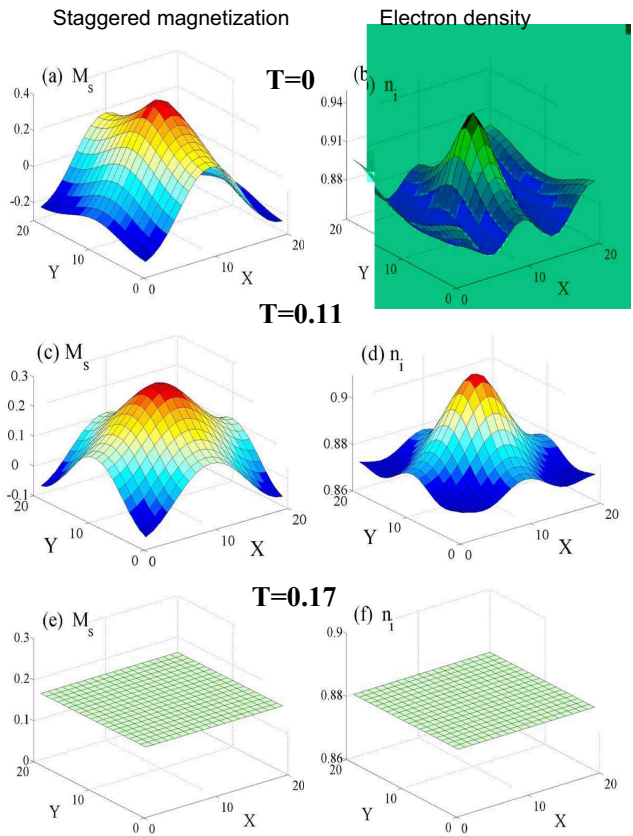}
\caption{\label{Fig4} Spatial variations of the local electron
density $n_i$ for various temperatures (a) T=0.0 (b) T=0.11 and
(c) T=0.17. The strength of the on-site repulsion $U=2.4$. The
averaged electron density is fixed at $\bar {n}=0.88$.}
\end{figure}
As discussed in our previous work~\cite{chen02}, the parameters
used to obtain Fig. 4 yield stripe like structures in the DSC, SDW
and CDW order parameters at very low temperature. In the present
work, the DSC and SDW orders coexist in underdoped sample below
$T_c$, which implies $T_{AF} > T_c$.  So the question arises: will
these stripe like structures  persist with increasing temperature?
From our calculations,  the spatial profiles of these order
parameters  may go through three different symmetries as $T$
increases.  At  $T=0$, the spatial distribution  of  SDW ($M_s$)
and CDW ($n_i$) orders  exhibit quasi-one-dimensional stripe like
behavior (see Figs. 4(a) and 4(b)) . When $T$ is raised to
$T=0.11$ above a characteristic temperature $T_{s}$ ($\sim 0.1$),
the symmetry of the patterns change to isotropic and
two-dimensional (see Figs. 4(c) and 4(d)).   Above $T_c$ (${\sim
0.16}$) and at $T=0.17$, the residual AF and electron density
become uniform as shown in Figs. 4(e) and 4(f). The AF order
vanishes when $ T > T_{AF}$.

Finally we display the temperature evolution of the vortex charge
number $Q_v/e$  for the underdoped case $\delta=0.12$ with $U=2.4$
and $U=2.2$ in Fig. 5.
\begin{figure}[t]
\includegraphics[width=7.2cm]{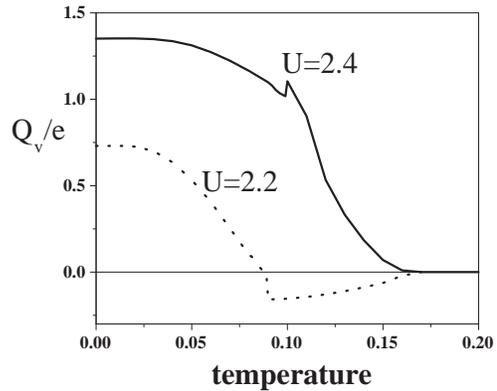}
\vspace{-0.4cm} \caption{\label{Fig5} The vortex charge number
$Q_v/e$ for $U=2.4$(solid line), $U=2.2$(dotted line) as a
function of temperature T. The averaged electron density is fixed
at $\bar {n}=0.88$.}
\end{figure}
Comparing with Fig. 3, the magnitudes of vortex charges for both
cases in Fig. 5 at low temperatures become larger as a result of
stronger AF order near the vortex core in  underdoped samples. For
$U=2.4$, the $Q_v/e$ versus temperature curve (see the solid line
in Fig. 5) exhibits a discontinuity at $T_{s} \sim 0.1$ which
reflects the characteristic temperature between two different
symmetries. For $U=2.2$,  the spatial distribution patterns for
SDW and CDW are always isotropic and two-dimensional with
$T_{AF}\sim 0.09$ where a appreciable discontinuity in $Q_v/e$
(see the dotted line in Fig. 5) is present.

Recently STM experiments~\cite{Pan,Lang} showed  the existence of
inhomogeneities in the electron density even for  the optimally
doped  Bi$_2$Sr$_2$CaCu$_2$O$_{8+x}$,  there  the hole density
varies from $0.1$ to $0.2$.  From our previous
paper~\cite{vcharge} and the present study, we found that there
may exist two types of vortex cores either negatively charged or
positively charged. By increasing $T$, the number of negatively
charged vortices decreases, while positively charged vortices
increase. This may have profound effect on the mixed state
transport properties in HTS. Finally we wish to point out that our
numerical study has been performed in a strong magnetic field. So
far we are not able to extend this calculation to a much weaker
field. However, according to the extrapolation
scenario~\cite{vcharge}, the magnitude of the induced AF order
follows roughly a linear relationship with $B$.  The strength of
the AF order and the magnitude of vortex charge are estimated  to
be much smaller as $B$ is reduced to the experimental value ($\sim
9.4 T$). Despite this difficulty, our numerical approach should
give a consistent description to the behavior of vortex charges in
HTS.

We are grateful to Prof. S.H. Pan, B. Friedman and Dr. J.X Zhu for
useful discussions. This work was supported by the Robert A. Welch
Foundation and by the Texas Center for Superconductivity at the
University of Houston through the State of Texas. ZDW also
appreciates  the support from the RGC grant of Hong Kong
(HKU7092/01P), the  Ministry of Science and Technology of China
under grant No. 1999064602, and the URC fund of HKU.

\end{document}